\documentclass[aps,pra,english,reprint,longbibliography,nofootinbib,floatfix]{revtex4-2}
\usepackage[utf8]{inputenc}
\usepackage[stretch=20]{microtype}
\usepackage{graphicx}
\usepackage[shortlabels]{enumitem}
\usepackage[draft=false, hidelinks]{hyperref}

\input{preamble_widebar}
\usepackage{nccmath}
\usepackage{mathtools} 
\usepackage{isomath}
\usepackage{relsize}
\usepackage{amssymb}
\usepackage{mathrsfs}
\usepackage{siunitx}
\usepackage{braket}
\usepackage{mleftright}
\mleftright

\newcommand*\conj[1]{\widebar{#1}}

\AtBeginDocument{}

\newcommand*\Laplace{\mathrm{\Delta}}
\newcommand{\Veff}{V_\text{ind}}

\AtBeginDocument{

}
\DeclarePairedDelimiter\abs{\lvert}{\rvert}

\newcommand*\B{\text{\textsmaller{B}}}
\newcommand*\IB{\text{\textsmaller{IB}}}
\newcommand*\BB{\text{\textsmaller{BB}}}

\newcommand*\mB{m_\B}
\newcommand*\VIB{V^{\IB}}
\newcommand*\VBB{V^{\BB}}
\newcommand*\aIB{a_\IB}
\newcommand*\aBB{a_\BB}

\newcommand*\nB{n}
\newcommand*\EGP{E_\text{\textsmaller{GP}}}
\newcommand*\ENLGP{E_\text{\textsmaller{NLGP}}}
\newcommand*\Eb{E_\text{b}}
\newcommand*\xIl{x_{\mathrm{I},1}}
\newcommand*\xIr{x_{\mathrm{I},2}}
\newcommand*\vol{\Omega}

\allowdisplaybreaks
\begin{document}

\title{Medium-induced Interaction Between Impurities in a Bose-Einstein Condensate}
\author{Moritz Drescher}
\author{Manfred Salmhofer}
\author{Tilman Enss}
\affiliation{Institut für Theoretische Physik, Universität Heidelberg,
  D-69120 Heidelberg, Germany}

\begin{abstract}
  We consider two heavy particles immersed in a Bose-Einstein condensate in three dimensions and compute their mutual interaction induced by excitations of the medium.
  For an ideal Bose gas, the induced interaction is Newtonian up to a shift in distance which depends on the coupling strength between impurities and Bosons.
  For a real BEC, we find that on short distances, the induced potential is dominated by three-body physics of a single Boson bound to the impurities, leading to an Efimov potential.
  At large distances of the order of the healing length, a Yukawa potential emerges instead.
  In particular, we find that both regimes are realized for all impurity-boson couplings and determine the corresponding crossover scales.
  The transition from the real to the ideal condensate at low gas parameters is investigated.
\end{abstract}

\maketitle

\section{Introduction}

Interactions between particles in fundamental physics emerge by local excitations of fields.
When the particles are sufficiently slow, the field's degrees of freedom can be integrated out, leading to an effective action-at-a-distance, which may be described by a distance-dependant potential.
The properties of the induced potential depend on properties of the field and become highly non-trivial when the field itself consists of interacting particles.

Recent progress in the creation and control of ultracold gas systems has enabled experiments where the influence of such interacting media of particles on impurities can be studied.
In the case where the medium is a Bose-Einstein condensate (BEC), the impurities are known as Bose polarons and have been realized experimentally \cite{Hu2016, Jorgensen2016, Yan2020, Skou2021} and studied in and out of equilibrium \cite{Rath2013, Shchadilova2016, Levinsen2017, Drescher2019, Drescher2020, Dzsotjan2020, Schmidt2022, Drescher2021, Levinsen2021, Rose2022, Enss2022}.

A number of theoretical works have studied interactions induced by a BEC.
\cite{Zinner2013} applies Bogoliubov theory to study how the Efimov hierarchy generated by a single Boson is modified if a whole condensate is present.
In \cite{Naidon2018}, a variational wave function with a single Bogoliubov excitation led to a Yukawa potential for weak impurity-boson (IB) coupling and to an Efimov potential near a resonance.
A perturbative treatment in momentum space was given in \cite{Camacho-Guardian2018},
a path-integral approach in \cite{Panochko2022}.
\cite{Camacho-Guardian2018a} investigates the formation of bipolarons.
\cite{Panochko2021} discusses the case of an ideal condensate, including finite temperature.
In \cite{Fujii2022}, general superfluids with a linear dispersion were considered, which, when applied to a BEC, treats the case of impurity separations large compared to the healing length.
It was shown that first-order effects in the gas parameter, which would lead to a Yukawa potential, are dominated at these scales by second-order effects and a van-der-Waals potential emerges.
\cite{Jager2022} consider a soft IB potential while in \cite{Ding2022, Astrakharchik2023}, the impurities are ionic and interact via an $r^{-4}$ potential and in \cite{Bighin2022a}, a spinor BEC is investigated.
Also related are works on 1d condensates \cite{Schecter2014, Dehkharghani2018, Charalambous2019, Mistakidis2020, Will2021, Petkovic2022} and on Fermi polarons \cite{Nishida2009, DeSalvo2019, Edri2020, Enss2020, Tran2021}.

\begin{figure}
    \includegraphics{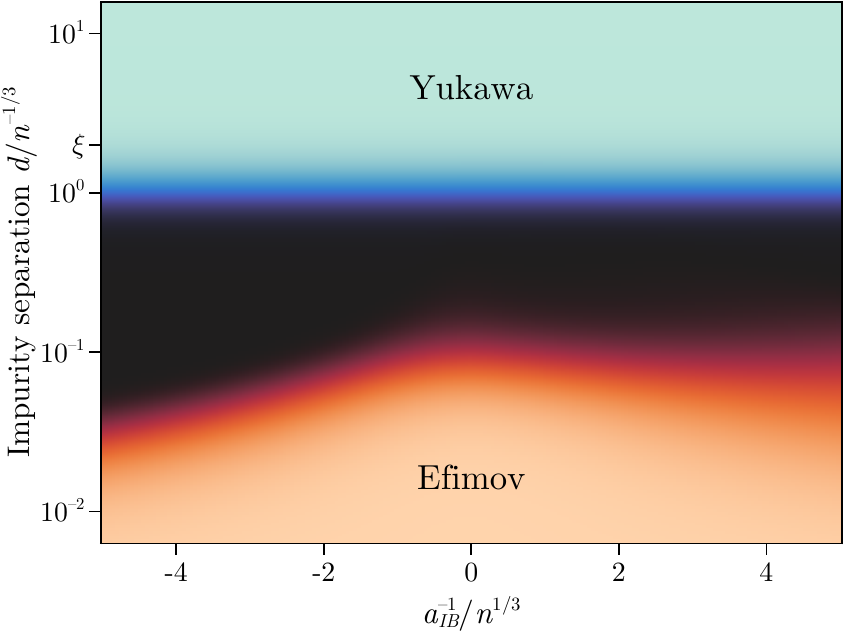}
    \caption{\label{fig:scaling_regimes}
    Sketch of the regimes of Efimov and Yukawa scaling of the induced potential at a gas parameter of $n\aBB^3 = 10^{-6}$.
    The color gradient indicates how well the induced potential matches the Efimov or Yukawa scaling forms
    (see Figs.\ \ref{fig:scaling_loglog}, \ref{fig:scaling_loglog_all}).
    At small impurity separations $d$, a three-body bound state is formed, which has an Efimov scaling $\Eb \sim d^{-2}$ for $d \ll \abs{a_{IB}}$ where $\aIB$ is the impurity-boson scattering length.
    At separations comparable to the healing length, the induced potential takes a Yukawa form $\sim \exp(-d\sqrt 2 / \xi)/d$ with the healing length $\xi = 1 / \sqrt{8\pi n \aBB}$.
    At even larger separations, a van-der-Waals scaling has been shown to emerge \cite{Fujii2022} (not shown in figure).}
\end{figure}

In this paper, we consider the experimentally relevant case of a short (though finite) range potential.
We employ Gross-Pitaevskii theory (GPT) as well as a recently developed non-local extension (NLGPT, \cite{Drescher2020}) to be able to treat large deformations of the condensate.
We find that the Efimov and Yukawa scaling of the induced potential, previously found at different coupling strengths \cite{Naidon2018}, in fact exist at every coupling but at different impurity separations as depicted in Fig.\ \ref{fig:scaling_regimes}.
For an ideal BEC, on the other hand, the potential is shown to be Newtonian up to a shift in distance.
Approaching the ideal gas case by a weakly interacting condensate, we show that the Newtonian potential emerges only at very weak gas parameters of order $10^{-15}$ in an intermediate distance regime.

\section{Model}

The Hamiltonian for two stationary impurities in a BEC is given by
\begin{multline} \label{eq:Hamiltonian}
  H = \int_x \, a^\dagger_x \left( -\frac{\Laplace}{2\mB} + \VIB_{x - \xIl}
  + \VIB_{x - \xIr} \right) a^{}_x \\
      + \int_{x,y} \, a^\dagger_x a^\dagger_y \VBB_{x - y} a^{}_x a^{}_y
\end{multline}
where $a^{(\dagger)}_x$ are the boson annihilation and creation operators at position $x \in \mathbb{R}^3$, $m_B$ the boson mass, $\VIB$ and $\VBB$ the impurity-boson and boson-boson
interaction potentials of scattering lengths $\aIB$ and $\aBB$, respectively, and $x_{\mathrm{I}, i}$ the
impurity positions. Their distance will be denoted $d = \abs{\xIr - \xIl}$.
We are looking for the ground state energy $E_d$ of this Hamiltonian at different impurity separations $d = \abs{\xIl - \xIr}$ to compare it with the energy of two individual impurities $E_\infty$ and thus obtain the medium-induced interaction
\begin{equation} \label{eq:Veff}
  \Veff(d) = E_d - E_\infty.
\end{equation}

\subsection{Basic properties}

Before turning to the methods applied to the Hamiltonian, let us recall some elementary analytical results, which
serve as a reference.

\paragraph{Single-Boson Problem.}
The three-body problem of a single Boson and two stationary impurities can be solved in the case of an IB contact potential.
Clearly, the effective potential formed by the two impurities is the stronger the closer the impurities are.
One finds, that a bound state exists whenever $\aIB^{-1} > -d^{-1}$ (compared to $\aIB^{-1} > 0$ for a single impurity) with binding energy of $\Eb = \kappa^2/2\mB$ where
\begin{equation} \label{eq:binding_energy}
    \kappa = \frac{1}{\aIB} + \frac{1}{d} W_0(e^{-d/\aIB})
\end{equation}
and $W_0$ is the principal branch of the Lambert W function, i.e.\ the inverse of $we^w$ defined on $[-1/e, \infty)$
\cite{Nishida2009, Enss2020}.
At short distances or resonant coupling $\aIB^{-1} = 0$, the binding energy takes the form $\Eb \sim d^{-2}$, which is reminiscent of Efimov physics and admits a tower of bound states.
We may expect to find the same behaviour at small distances in the many-body problem \cite{Sun2017}, as a single boson gets bound to the impurities and repels the other bosons, c.f.\ \cite{Levinsen2021}.

\paragraph{Ideal BEC.}
In absence of BB interactions, the ground state of \eqref{eq:Hamiltonian} is a product state with all bosons
being in the ground state of the single-boson problem. If this problem allows for a bound state,
the many-body ground state energy will be $-\infty$ as all bosons enter the bound state and the BEC collapses.
Otherwise, all bosons are in the zero-mode, i.e.\ the lowest scattering state, and the ground state energy will be finite \cite{Drescher2021}.
It is not zero, however, because in finite volume $\vol$, the lowest scattering state has an energy of order $1/\vol$, which, when occupied $N$ times, does not vanish in the thermodynamic limit.

Again, the problem can be solved explicitly for contact interactions.
Regarding the two impurities as forming a single effective potential of scattering length $a_\text{eff}$,
the ground state energy is given by $E = 4\pi a_\text{eff} n/2\mB$ (c.f.\ \cite{Drescher2021}).
The zero mode $\phi^0$ defines $a_\text{eff}$ via
$\phi^0(x) \rightarrow 1 - a_\text{eff} / \abs{x}$ as $\abs{x} \rightarrow \infty$.
It reads%
\footnote{Indeed, one easily checks that this formula fulfils
\begin{equation*}
  \left(-\frac{\Laplace}{2\mB} 
  + \VIB(x - \xIl) + \VIB(x - \xIr) \right) \phi(x) = 0
\end{equation*}
where
$\VIB(\abs{r}) \phi(r) = \frac{4\pi \aIB}{2\mB} \delta^3(r) \frac{\partial}{\partial \abs{r}} \abs{r} \phi(\abs{r}e_r)$
is the contact potential of scattering length $\aIB$.
}
\begin{equation*}
  \phi_x^0 = 1 - \frac{\alpha}{\abs{x - \xIl}} - \frac{\alpha}{\abs{x - \xIr}}
\end{equation*}
where $\alpha = 1 / (\aIB^{-1} + d^{-1})$, so we obtain
$a_\text{eff} = 2\alpha$ and
\begin{equation*}
  E_d = \frac{8\pi n / 2\mB }{\aIB^{-1} + d^{-1}}.
\end{equation*}
This formula has already been derived by a different method in \cite{Panochko2021} and leads to a “shifted Newtonian”
induced attractive potential
\begin{equation*}
  \Veff(d) = E_d - E_\infty = -\frac{8\pi \aIB^2 n}{2\mB} \frac{1}{\aIB + d}
\end{equation*}
whenever $a_\text{eff} < 0$, i.e.\ $\aIB < 0$ and $\abs{\aIB} < d$ (otherwise, $\Veff = -\infty$ because a bound state exists which can be entered by all bosons).

Note that the regions of applicability of the single-boson analysis vs.\ the ideal BEC analysis are complementary: the former requires a bound state to have any effect, while the latter requires absence of a bound state lest the BEC collapses.
It is for this reaseon that the Newtonian potential does not reproduce the Efimov effect: it is applicable only at $d > \abs{\aIB}, \aIB < 0$, while Efimov scaling occurs at $d < \abs{\aIB}$.

\subsection{Methods}

For the full interacting problem, we employ Gross-Pitaevskii theory (GPT) as well as its non-local extension (NLGPT) developed
recently in \cite{Drescher2020}.

\paragraph*{GPT.}
If the boson density varies on a length scale larger than the BB interaction range, a 
local density approximation can be applied, and the zero temperature condensate can be described by a single condensate wave
function $\phi$. Using that the ground state energy density of a BEC is $4\pi \aBB \nB^2 / 2\mB$, this leads to the well-known
Gross-Pitaevskii energy functional
\begin{multline} \label{eq:gp_functional}
  \EGP = \int_x \conj{\phi_x} \left( -\frac{\Laplace}{2\mB}  + \VIB_{x - \xIl} 
    + \VIB_{x - \xIr} \right) \phi^{}_x \\
    + \frac{4\pi \aBB}{2\mB} \int_x \abs{\phi_x}^4.
\end{multline}
Most results of this work are obtained by minimizing this functional. To justify the local
density approximation, the IB potentials must not be too sharp \cite{Drescher2020}: For instance for a
contact potential, $\phi$ would have poles $1/\abs{x - x_{\mathrm{I}, i}}$ and the BB term would
diverge. At a gas parameter of $n\aBB^3 = 10^{-6}$, we employ a Gaussian potential $\VIB(r) \sim \exp(-r^2/\sigma^2)$ of range $\sigma = 0.01n^{-1/3}$.
Since then $\sigma = \aBB$ is not small compared to $\aBB$, this is, in principle, not soft enough for GPT to be applicable and indeed, we observe differences in energy compared to NLGPT with the same parameters.
However, this error is almost constant in $d$, such that GPT yields the correct result for the induced potential \eqref{eq:Veff} for large separations even for this rather small value of $\sigma$.

\paragraph*{NLGPT.}
For the case of strong local variation of $n$, we have shown in \cite{Drescher2020}
how to extend GPT to a non-local theory by not integrating out the local BB correlations. For two static
impurities, the resulting energy functional reads
\begin{multline} \label{eq:nlgp_functional}
  \ENLGP = \int_x \conj{\phi_x} \left( -\frac{\Laplace}{2\mB}  + \VIB_{x - \xIl} 
    + \VIB_{x - \xIr} \right) \phi^{}_x \\
    + \int_{x, y} \abs{\phi_x}^2 \abs{\phi_y}^2 
      g_{x-y}.
\end{multline}
Here, $g = \frac{1}{2} \left[ \frac{\abs{\nabla f}^2}{\mB} + \VBB \abs{f}^2 \right]$ where $f$ is the zero-energy solution to the two-boson problem:
$\bigl(-\frac{\Laplace}{\mB} + \VBB \bigr)f = 0$.
We will employ a hard sphere potential, for which $f(x) = \max(0, 1 - \aBB/\abs{x})$.
Note that $g$ is an approximate delta function of range $\aBB$ and height $\int g = 4\pi \aBB / 2\mB$ and if $\phi$ is almost constant on these scales, the GP functional is recovered.
We compute individual data points by minimizing $\ENLGP$ with the same IB interaction potential $\VIB$ as 
for GPT to assert the validity of the latter at large distances as well as to obtain additional results for smaller distances.
Still, we limit ourselves to $d \gtrsim \sigma$ where the resulting scaling of $\Veff$ is universal.

\section{Results}

\subsection{Scaling behaviour of the induced potential}

\begin{figure}
    \includegraphics{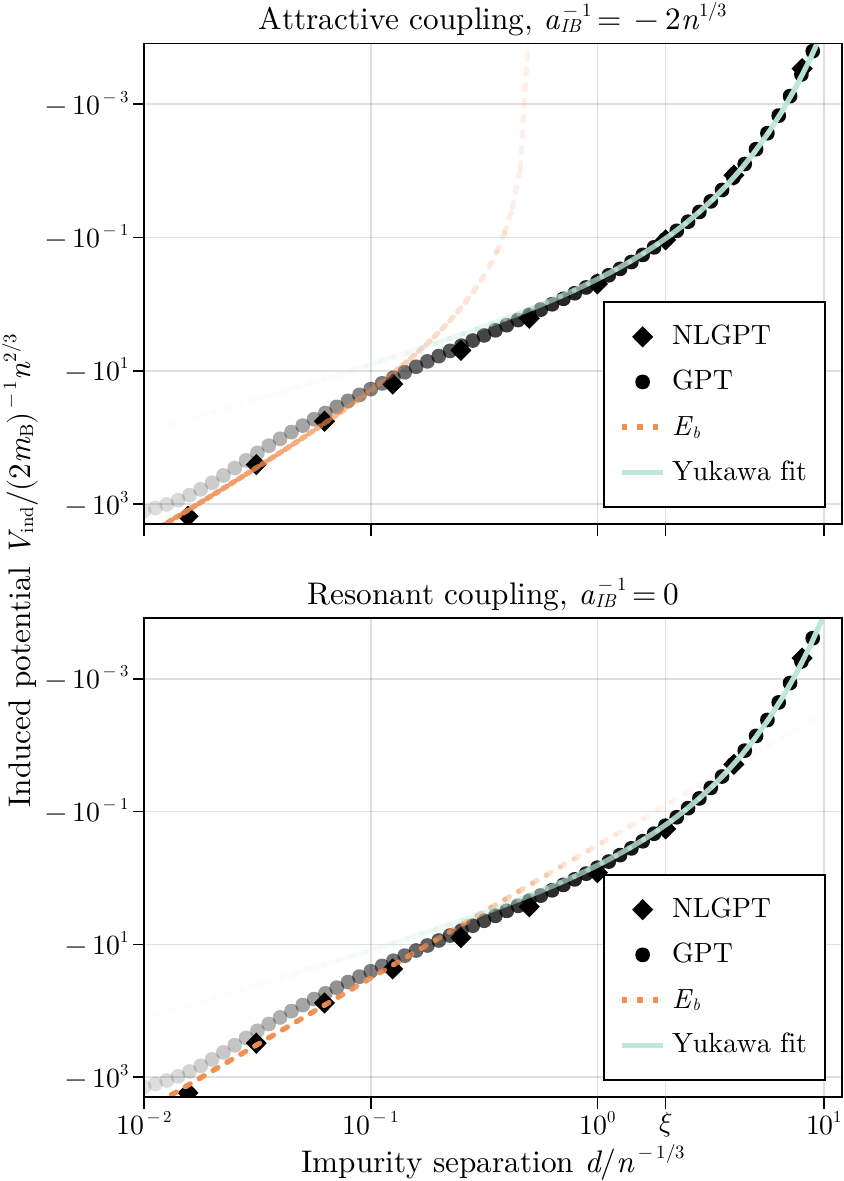}
    \caption{\label{fig:scaling_loglog}
    Induced potential at a gas parameter of $n\aBB^3 = 10^{-6}$ and two different coupling strengths (more can be found in the appendix).
    The circles are the results of the GP simulation and have a Yukawa scaling at large impurity separation.
    The non-local GPT (diamonds) confirms the applicability of GPT in this regime and remains valid at shorter separations, where GPT results are off by a factor of about 1.5.
    Here, the three-body binding energy $\Eb$ is approached, which exhibits Efimov scaling $\sim d^{-2}$ for $d \ll \abs{\aIB}$.
    }
\end{figure}

Minimizing the GP and NLGP energy functionals at different impurity separations $d$, we obtain
the induced potential shown in Fig.\ \ref{fig:scaling_loglog}.
It exhibits the following scaling behaviour:
\begin{enumerate}[(a)]
\item Large-distance behaviour: Yukawa regime.
  At large distances, the induced
  potential falls off like a Yukawa potential \cite{Naidon2018, Jager2022}
  \begin{equation*}
    \Veff(d) \sim \frac{e^{-d\sqrt 2 / \xi}}{d}.
  \end{equation*}
  In GPT, this can be seen as follows: The energy of a single impurity in a flat BEC is proportional to $n$.
  At large impurity separation, each impurity feels an almost flat density modulation caused by the other
  impurity, so we should have $E_d \sim n(d)$ where $n(d)$ is the density profile of a BEC with one impurity
  at zero while $n_0$ denotes the unperturbed density.
  $n(d)$ is obtained from the zero-energy GPE with one impurity: At distances larger than the potential range, it reads (see also \cite{Massignan2021})
  \begin{align}
    0 &= -\frac{\Laplace}{2\mB} \phi + \frac{8\pi\aBB}{2\mB} \phi (\phi^2 - n_0) \notag\\
    \Rightarrow \quad 0 &= -\delta \phi'' - \frac{2}{d} \delta\phi' +\frac{2}{\xi^2} \delta\phi + \mathscr{O}(\delta\phi^2) \label{eq:gpe_for_deltaphi}
  \end{align}
  where we have set $\phi(d) = \sqrt n_0 + \delta \phi(d)$.
  This is solved by $\delta\phi \sim e^{-d\sqrt 2 / \xi} / d$ and $\Veff(d) \sim n(d) - n_0 \simeq 2\delta\phi$.
  The linear approximation in $\delta \phi$ is valid when $\delta \phi$ is small, which is typically the case for distances on the order of $\xi$, except at weak coupling, in which case distances large compared to $\aIB$ are sufficient.
  Note that this argument does not require GPT to be valid close to an impurity, as it is only used to determine the density profile far away from one impurity.

  It must be noted though that the assumption of a finite-range impurity-boson potential was crucial for the
  derivation: if $\VIB$ has power-law tails, an additional term $2\mB \VIB \sqrt n$ appears in 
  \eqref{eq:gpe_for_deltaphi}. At radii large compared to $\xi$, the asymptotic solution is then
  $\delta \phi \simeq -\xi^2 \mB \sqrt{n} \VIB$, such that $\Veff(d) \sim \VIB(d)$.
  This effect can bee observed in Ref.\ \cite{Ding2022} where ions in a BEC interact via an $r^{-4}$
  potential, resulting in an $r^{-4}$ tail in the induced interaction%
  \footnote{
      The above analysis with the weak-coupling single-impurity energy $E(n) = 4\pi \aIB n / 2\mB$ actually leads to a slightly
      different prefactor $\Veff(d) = - (\aIB/\aBB) \VIB(d)$ than
      in Ref.\ \cite{Ding2022}, where $\aIB$ is replaced by its Born approximation.
  }.
  Also, since the analysis is based on Gross-Pitaevskii theory, only the leading order in the gas parameter is captured.
  At distances large compared to the healing length, sub-leading terms eventually become dominant as the Yukawa potential vanishes \cite{Fujii2022}.

\item Short-distance behaviour: Efimov regime. 
  At small distances, the induced potential approaches the three-body binding energy $\Eb$.
  This indicates that a single Boson gets bound to the impurity and blocks the bound state for the others due to its repulsion as described in \cite{Levinsen2021}.
  When $d$ is small against $\aIB$, $\Eb$ and thus also $\Veff$ scale like $d^{-2}$ as in the Efimov effect.
  At resonance (second panel in Fig.\ \ref{fig:scaling_loglog}), $\Eb$ is directly proportional to $d^{-2}$ even for larger distances.
  Nonetheless, we find that the Efimov regime in the many-body system has a finite range of the order of the mean particle distance.
  This is natural since the picture of one molecule surrounded by a bath breaks down as the size of the Efimov trimer increases and extends into the bath \cite{Sun2017}.
\end{enumerate}

We are not interested here in even shorter ranges of $d < \sigma$, where the scaling of the potential becomes non-universal and 
starts to depend on the shape of $\VIB$.

Comparing to the results of \cite{Naidon2018}, we thus also find regions of Yukawa and Efimov scaling, but they are not delimited
by the scattering length alone.
Instead, we always find both regimes but for different impurity separations while the scattering length controls the size of these regimes.
A sketch of this is shown in Fig.\ \ref{fig:scaling_regimes}.
The length scale that delimits the Efimov regime is, for weak coupling, the scattering length $\aIB$ \cite{Zinner2013}, since the three-body binding energy scales like $d^{-2}$ only for $d \ll \abs{\aIB}$, see \eqref{eq:binding_energy}.
If the impurity separation becomes comparable to the mean boson distance $\sim n^{-1/3}$, the picture of an individual boson mediating the interaction no longer applies.
Thus, $d \ll n^{-1/3}$ is another condition delimiting the Efimov region and it becomes the dominant one near the resonance.
The Yukawa regime, on the other hand, starts when the local density variation $\delta \phi$ from the treatment above becomes sufficiently small for the linear approximation in \eqref{eq:gpe_for_deltaphi} to be applicable.
This is the case when $d$ is of the order of the healing length $\xi$, or, if $0 < -\aIB \ll \xi$, when $d \gg \abs{\aIB}$.
More quantitatively, we find good agreement for $d \gtrapprox \xi / 2$ or $d \gtrapprox 20\abs{\aIB}$.
The first condition dominates in the parameter region considered in Fig.\ \ref{fig:scaling_regimes}.
In the appendix, we show results at $\aIB = -0.01 n^{-1/3}$ exhibiting a larger Yukawa regime.

We note that when $d$ is large compared to $\xi$, \cite{Fujii2022} finds that the quickly decaying Yukawa potential becomes dominated by second-order effects, which lead to van-der-Waals scaling.

\subsection{Approach to ideal BEC}

\begin{figure}
  \includegraphics[width=\linewidth]{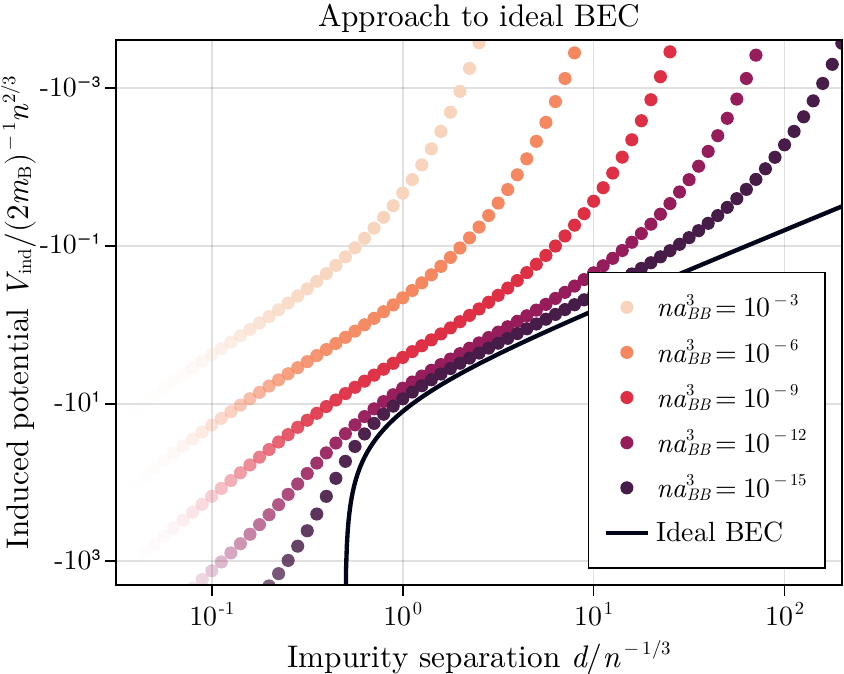}
    \caption{\label{fig:approach_to_ideal}Induced potential obtained with GPT for varying gas parameters at fixed impurity coupling $\aIB = -0.5n^{-1/3}$.
    The analytical curve for the ideal BEC exhibiting Newtonian behaviour is approached at very low gas parameters in an intermediate distance regime $\abs{\aIB} < d < \xi$.}
\end{figure}

In the beginning, we have shown that attractively coupled impurities in an ideal BEC lead to a shifted Newtonian potential.
This matches neither the Efimov nor the Yukawa potential found for the real BEC, so one may wonder how the Newtonian regime
emerges as an ideal BEC is approached.
In Fig.\ \ref{fig:approach_to_ideal}, we solve the GPE at fixed impurity coupling $\aIB$ for a sequence of BEC gas parameters $n\aBB^3$.
As it is lowered, the ideal gas result is approached in an intermediate distance regime, which is bounded by $\abs{\aIB}$ from below and $\xi$ from above.
However, the approach is very slow and gas parameters of the order of $10^{-15}$ are required for it to be visible.
This is a consequence of the slow splitting of length scales in a BEC: the healing length in units of $n^{-1/3}$ scales with
order $-1/6$ in the gas parameter, so very small values of the latter are required to push the Yukawa regime to sufficiently large distances to make room for an intermediate Newtonian scaling regime.

\section{Conclusion}

We have investigated the problem of a force between particles induced by an interacting Bose-Einstein condensed medium by minimizing local and non-local Gross-Pitaevskii energy functionals.
We found that both the Efimov and Yukawa scaling of the mediated potential can be observed in certain distance regions with an intermediate cross-over regime (Fig.\ \ref{fig:scaling_regimes}).
This is in contrast to previous studies predicting either one or the other depending on the coupling strength.
Our result, valid for impurity separations up to the order of the healing length, complements a recent study treating separations larger than the healing length \cite{Fujii2022}.
A non-interacting condensate, on the other hand, was shown to induce a shifted Newtonian potential.
The cross-over from interacting to non-interacting condensate was found to take place at low gas parameters of order $10^{-15}$,
where the Newtonian behaviour emerges in an intermediate-distance regime.
Experimentally, induced interactions are expected to lead to an energy shift depending on impurity concentration, which may be probed by radio-frequency spectroscopy \cite{Hu2016, Jorgensen2016, Yan2020, Skou2021}.
To investigate the distance dependence, the impurities can be confined to individual microtraps \cite{Ding2022}.
For future work, it will be interesting to investigate the influence of finite impurity mass in a region where it is still large compared to the Boson mass, such that the picture of mediated forces can be applied.

\acknowledgements

The authors acknowledge interesting discussions with K.\ Fujii, R.\ Schmidt, M.\ Weidemüller, E.\ Lippi, M.\ Rautenberg and F.\ Rose.
This work is funded by the DFG (German Research Foundation) under Project-ID 273811115, SFB 1225 ISOQUANT
and under Germany's Excellence Strategy EXC2181/1-390900948 (the Heidelberg STRUCTURES Excellence Cluster).

\appendix

\section{Induced potential for more values of the coupling}

\begin{figure*}[h]
  \includegraphics[height=0.95\textheight]{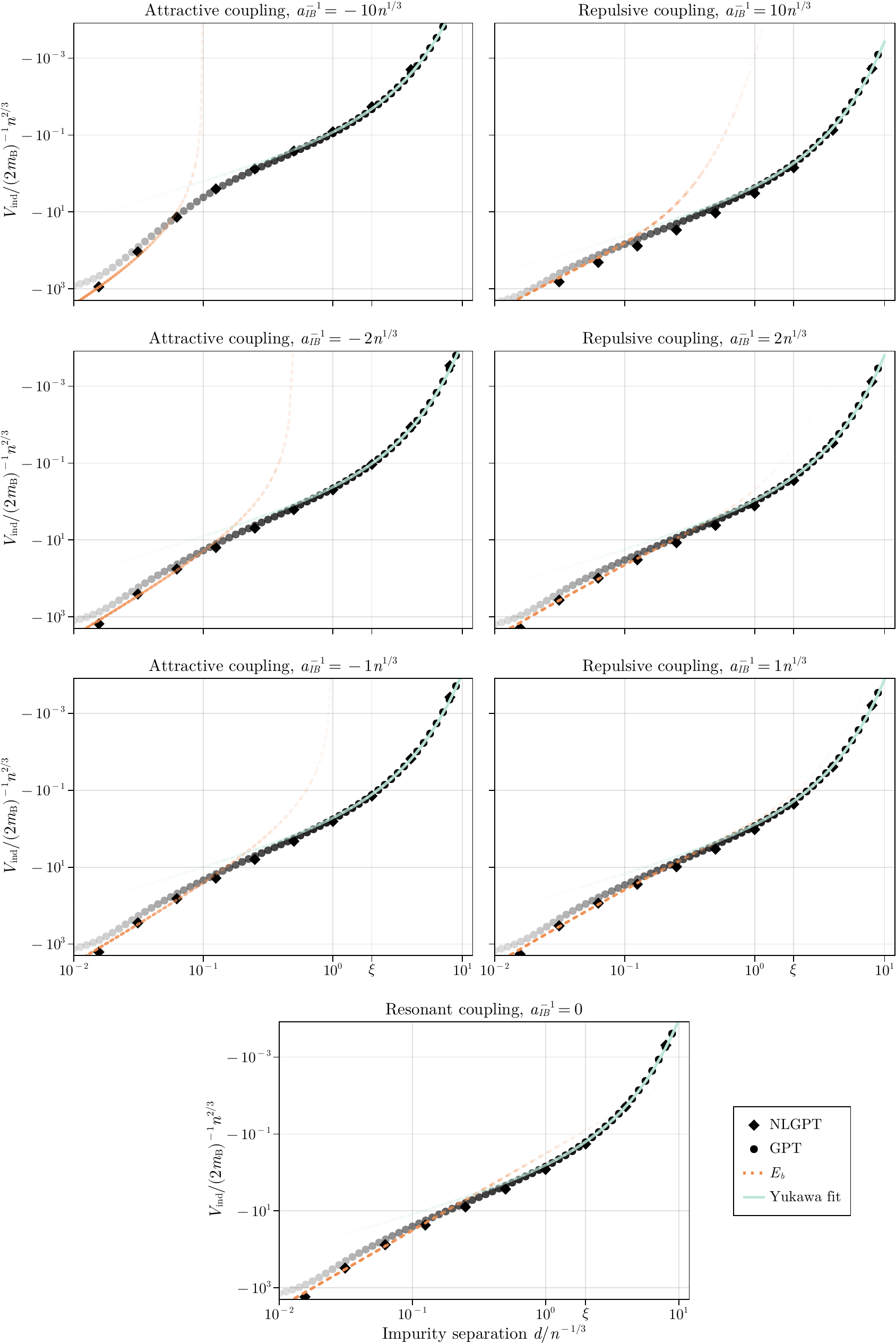}
  \caption{\label{fig:scaling_loglog_all}
  Induced potential for various coupling strengths $\aIB$.
  The gas parameter is $n\aBB^3 = 10^{-6}$ as in Fig.\ \ref{fig:scaling_loglog}.
  }
\end{figure*}

\begin{figure}[h]
    \includegraphics{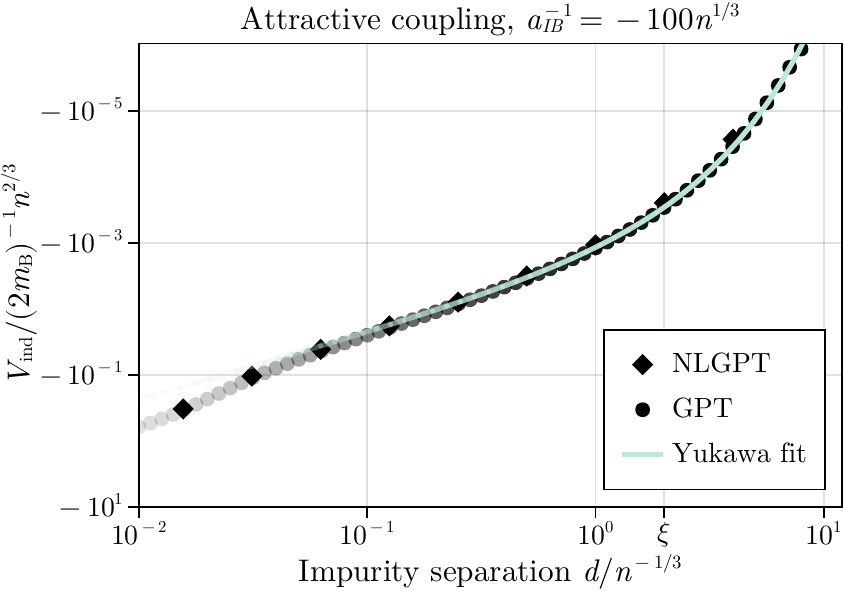}
    \caption{\label{fig:scaling_loglog_weak}
    At very weak coupling, the Yukawa region is extended to $d \gtrapprox 20\abs{\aIB}$.
    }
\end{figure}

In Fig.\ \ref{fig:scaling_loglog_all}, we show the induced potentials as in Fig.\ \ref{fig:scaling_loglog} for more values of the coupling $\aIB$, including the repulsive side of a resonance.
The scaling regimes observed from these curves lead to Fig.\ \ref{fig:scaling_regimes}.
Fig.\ \ref{fig:scaling_loglog_weak} shows the growth of the Yukawa regime at very weak coupling.

\begin{thebibliography}{39}%
\makeatletter
\providecommand \@ifxundefined [1]{%
 \@ifx{#1\undefined}
}%
\providecommand \@ifnum [1]{%
 \ifnum #1\expandafter \@firstoftwo
 \else \expandafter \@secondoftwo
 \fi
}%
\providecommand \@ifx [1]{%
 \ifx #1\expandafter \@firstoftwo
 \else \expandafter \@secondoftwo
 \fi
}%
\providecommand \natexlab [1]{#1}%
\providecommand \enquote  [1]{``#1''}%
\providecommand \bibnamefont  [1]{#1}%
\providecommand \bibfnamefont [1]{#1}%
\providecommand \citenamefont [1]{#1}%
\providecommand \href@noop [0]{\@secondoftwo}%
\providecommand \href [0]{\begingroup \@sanitize@url \@href}%
\providecommand \@href[1]{\@@startlink{#1}\@@href}%
\providecommand \@@href[1]{\endgroup#1\@@endlink}%
\providecommand \@sanitize@url [0]{\catcode `\\12\catcode `\$12\catcode
  `\&12\catcode `\#12\catcode `\^12\catcode `\_12\catcode `\%12\relax}%
\providecommand \@@startlink[1]{}%
\providecommand \@@endlink[0]{}%
\providecommand \url  [0]{\begingroup\@sanitize@url \@url }%
\providecommand \@url [1]{\endgroup\@href {#1}{\urlprefix }}%
\providecommand \urlprefix  [0]{URL }%
\providecommand \Eprint [0]{\href }%
\providecommand \doibase [0]{https://doi.org/}%
\providecommand \selectlanguage [0]{\@gobble}%
\providecommand \bibinfo  [0]{\@secondoftwo}%
\providecommand \bibfield  [0]{\@secondoftwo}%
\providecommand \translation [1]{[#1]}%
\providecommand \BibitemOpen [0]{}%
\providecommand \bibitemStop [0]{}%
\providecommand \bibitemNoStop [0]{.\EOS\space}%
\providecommand \EOS [0]{\spacefactor3000\relax}%
\providecommand \BibitemShut  [1]{\csname bibitem#1\endcsname}%
\let\auto@bib@innerbib\@empty
\bibitem [{\citenamefont {Hu}\ \emph {et~al.}(2016)\citenamefont {Hu},
  \citenamefont {{Van de Graaff}}, \citenamefont {Kedar}, \citenamefont
  {Corson}, \citenamefont {Cornell},\ and\ \citenamefont {Jin}}]{Hu2016}%
  \BibitemOpen
  \bibfield  {author} {\bibinfo {author} {\bibfnamefont {M.-G.}\ \bibnamefont
  {Hu}}, \bibinfo {author} {\bibfnamefont {M.~J.}\ \bibnamefont {{Van de
  Graaff}}}, \bibinfo {author} {\bibfnamefont {D.}~\bibnamefont {Kedar}},
  \bibinfo {author} {\bibfnamefont {J.~P.}\ \bibnamefont {Corson}}, \bibinfo
  {author} {\bibfnamefont {E.~A.}\ \bibnamefont {Cornell}},\ and\ \bibinfo
  {author} {\bibfnamefont {D.~S.}\ \bibnamefont {Jin}},\ }\bibfield  {title}
  {\bibinfo {title} {Bose {{Polarons}} in the {{Strongly Interacting
  Regime}}},\ }\href {https://doi.org/10.1103/PhysRevLett.117.055301}
  {\bibfield  {journal} {\bibinfo  {journal} {Physical Review Letters}\
  }\textbf {\bibinfo {volume} {117}},\ \bibinfo {pages} {055301} (\bibinfo
  {year} {2016})}\BibitemShut {NoStop}%
\bibitem [{\citenamefont {J{\o}rgensen}\ \emph {et~al.}(2016)\citenamefont
  {J{\o}rgensen}, \citenamefont {Wacker}, \citenamefont {Skalmstang},
  \citenamefont {Parish}, \citenamefont {Levinsen}, \citenamefont
  {Christensen}, \citenamefont {Bruun},\ and\ \citenamefont
  {Arlt}}]{Jorgensen2016}%
  \BibitemOpen
  \bibfield  {author} {\bibinfo {author} {\bibfnamefont {N.~B.}\ \bibnamefont
  {J{\o}rgensen}}, \bibinfo {author} {\bibfnamefont {L.}~\bibnamefont
  {Wacker}}, \bibinfo {author} {\bibfnamefont {K.~T.}\ \bibnamefont
  {Skalmstang}}, \bibinfo {author} {\bibfnamefont {M.~M.}\ \bibnamefont
  {Parish}}, \bibinfo {author} {\bibfnamefont {J.}~\bibnamefont {Levinsen}},
  \bibinfo {author} {\bibfnamefont {R.~S.}\ \bibnamefont {Christensen}},
  \bibinfo {author} {\bibfnamefont {G.~M.}\ \bibnamefont {Bruun}},\ and\
  \bibinfo {author} {\bibfnamefont {J.~J.}\ \bibnamefont {Arlt}},\ }\bibfield
  {title} {\bibinfo {title} {Observation of {{Attractive}} and {{Repulsive
  Polarons}} in a {{Bose-Einstein Condensate}}},\ }\href
  {https://doi.org/10.1103/PhysRevLett.117.055302} {\bibfield  {journal}
  {\bibinfo  {journal} {Physical Review Letters}\ }\textbf {\bibinfo {volume}
  {117}},\ \bibinfo {pages} {055302} (\bibinfo {year} {2016})}\BibitemShut
  {NoStop}%
\bibitem [{\citenamefont {Yan}\ \emph {et~al.}(2020)\citenamefont {Yan},
  \citenamefont {Ni}, \citenamefont {Robens},\ and\ \citenamefont
  {Zwierlein}}]{Yan2020}%
  \BibitemOpen
  \bibfield  {author} {\bibinfo {author} {\bibfnamefont {Z.~Z.}\ \bibnamefont
  {Yan}}, \bibinfo {author} {\bibfnamefont {Y.}~\bibnamefont {Ni}}, \bibinfo
  {author} {\bibfnamefont {C.}~\bibnamefont {Robens}},\ and\ \bibinfo {author}
  {\bibfnamefont {M.~W.}\ \bibnamefont {Zwierlein}},\ }\bibfield  {title}
  {\bibinfo {title} {Bose polarons near quantum criticality},\ }\href
  {https://doi.org/10.1126/science.aax5850} {\bibfield  {journal} {\bibinfo
  {journal} {Science}\ }\textbf {\bibinfo {volume} {368}},\ \bibinfo {pages}
  {190} (\bibinfo {year} {2020})}\BibitemShut {NoStop}%
\bibitem [{\citenamefont {Skou}\ \emph {et~al.}(2021)\citenamefont {Skou},
  \citenamefont {Skov}, \citenamefont {J{\o}rgensen}, \citenamefont {Nielsen},
  \citenamefont {{Camacho-Guardian}}, \citenamefont {Pohl}, \citenamefont
  {Bruun},\ and\ \citenamefont {Arlt}}]{Skou2021}%
  \BibitemOpen
  \bibfield  {author} {\bibinfo {author} {\bibfnamefont {M.~G.}\ \bibnamefont
  {Skou}}, \bibinfo {author} {\bibfnamefont {T.~G.}\ \bibnamefont {Skov}},
  \bibinfo {author} {\bibfnamefont {N.~B.}\ \bibnamefont {J{\o}rgensen}},
  \bibinfo {author} {\bibfnamefont {K.~K.}\ \bibnamefont {Nielsen}}, \bibinfo
  {author} {\bibfnamefont {A.}~\bibnamefont {{Camacho-Guardian}}}, \bibinfo
  {author} {\bibfnamefont {T.}~\bibnamefont {Pohl}}, \bibinfo {author}
  {\bibfnamefont {G.~M.}\ \bibnamefont {Bruun}},\ and\ \bibinfo {author}
  {\bibfnamefont {J.~J.}\ \bibnamefont {Arlt}},\ }\bibfield  {title} {\bibinfo
  {title} {Non-equilibrium quantum dynamics and formation of the {{Bose}}
  polaron},\ }\href {https://doi.org/10.1038/s41567-021-01184-5} {\bibfield
  {journal} {\bibinfo  {journal} {Nature Physics}\ ,\ \bibinfo {pages} {1}}
  (\bibinfo {year} {2021})}\BibitemShut {NoStop}%
\bibitem [{\citenamefont {Rath}\ and\ \citenamefont
  {Schmidt}(2013)}]{Rath2013}%
  \BibitemOpen
  \bibfield  {author} {\bibinfo {author} {\bibfnamefont {S.~P.}\ \bibnamefont
  {Rath}}\ and\ \bibinfo {author} {\bibfnamefont {R.}~\bibnamefont {Schmidt}},\
  }\bibfield  {title} {\bibinfo {title} {Field-theoretical study of the
  {{Bose}} polaron},\ }\href {https://doi.org/10.1103/PhysRevA.88.053632}
  {\bibfield  {journal} {\bibinfo  {journal} {Physical Review A}\ }\textbf
  {\bibinfo {volume} {88}},\ \bibinfo {pages} {053632} (\bibinfo {year}
  {2013})}\BibitemShut {NoStop}%
\bibitem [{\citenamefont {Shchadilova}\ \emph {et~al.}(2016)\citenamefont
  {Shchadilova}, \citenamefont {Grusdt}, \citenamefont {Rubtsov},\ and\
  \citenamefont {Demler}}]{Shchadilova2016}%
  \BibitemOpen
  \bibfield  {author} {\bibinfo {author} {\bibfnamefont {Y.~E.}\ \bibnamefont
  {Shchadilova}}, \bibinfo {author} {\bibfnamefont {F.}~\bibnamefont {Grusdt}},
  \bibinfo {author} {\bibfnamefont {A.~N.}\ \bibnamefont {Rubtsov}},\ and\
  \bibinfo {author} {\bibfnamefont {E.}~\bibnamefont {Demler}},\ }\bibfield
  {title} {\bibinfo {title} {Polaronic mass renormalization of impurities in
  {{Bose-Einstein}} condensates: {{Correlated Gaussian-wave-function}}
  approach},\ }\href {https://doi.org/10.1103/PhysRevA.93.043606} {\bibfield
  {journal} {\bibinfo  {journal} {Physical Review A}\ }\textbf {\bibinfo
  {volume} {93}},\ \bibinfo {pages} {043606} (\bibinfo {year}
  {2016})}\BibitemShut {NoStop}%
\bibitem [{\citenamefont {Levinsen}\ \emph {et~al.}(2017)\citenamefont
  {Levinsen}, \citenamefont {Parish}, \citenamefont {Christensen},
  \citenamefont {Arlt},\ and\ \citenamefont {Bruun}}]{Levinsen2017}%
  \BibitemOpen
  \bibfield  {author} {\bibinfo {author} {\bibfnamefont {J.}~\bibnamefont
  {Levinsen}}, \bibinfo {author} {\bibfnamefont {M.~M.}\ \bibnamefont
  {Parish}}, \bibinfo {author} {\bibfnamefont {R.~S.}\ \bibnamefont
  {Christensen}}, \bibinfo {author} {\bibfnamefont {J.~J.}\ \bibnamefont
  {Arlt}},\ and\ \bibinfo {author} {\bibfnamefont {G.~M.}\ \bibnamefont
  {Bruun}},\ }\bibfield  {title} {\bibinfo {title} {Finite-temperature behavior
  of the {{Bose}} polaron},\ }\href
  {https://doi.org/10.1103/PhysRevA.96.063622} {\bibfield  {journal} {\bibinfo
  {journal} {Physical Review A}\ }\textbf {\bibinfo {volume} {96}},\ \bibinfo
  {pages} {063622} (\bibinfo {year} {2017})}\BibitemShut {NoStop}%
\bibitem [{\citenamefont {Drescher}\ \emph {et~al.}(2019)\citenamefont
  {Drescher}, \citenamefont {Salmhofer},\ and\ \citenamefont
  {Enss}}]{Drescher2019}%
  \BibitemOpen
  \bibfield  {author} {\bibinfo {author} {\bibfnamefont {M.}~\bibnamefont
  {Drescher}}, \bibinfo {author} {\bibfnamefont {M.}~\bibnamefont
  {Salmhofer}},\ and\ \bibinfo {author} {\bibfnamefont {T.}~\bibnamefont
  {Enss}},\ }\bibfield  {title} {\bibinfo {title} {Real-space dynamics of
  attractive and repulsive polarons in {{Bose-Einstein}} condensates},\ }\href
  {https://doi.org/10.1103/PhysRevA.99.023601} {\bibfield  {journal} {\bibinfo
  {journal} {Physical Review A}\ }\textbf {\bibinfo {volume} {99}},\ \bibinfo
  {pages} {023601} (\bibinfo {year} {2019})}\BibitemShut {NoStop}%
\bibitem [{\citenamefont {Drescher}\ \emph {et~al.}(2020)\citenamefont
  {Drescher}, \citenamefont {Salmhofer},\ and\ \citenamefont
  {Enss}}]{Drescher2020}%
  \BibitemOpen
  \bibfield  {author} {\bibinfo {author} {\bibfnamefont {M.}~\bibnamefont
  {Drescher}}, \bibinfo {author} {\bibfnamefont {M.}~\bibnamefont
  {Salmhofer}},\ and\ \bibinfo {author} {\bibfnamefont {T.}~\bibnamefont
  {Enss}},\ }\bibfield  {title} {\bibinfo {title} {Theory of a resonantly
  interacting impurity in a {{Bose-Einstein}} condensate},\ }\href
  {https://doi.org/10.1103/PhysRevResearch.2.032011} {\bibfield  {journal}
  {\bibinfo  {journal} {Physical Review Research}\ }\textbf {\bibinfo {volume}
  {2}},\ \bibinfo {pages} {032011(R)} (\bibinfo {year} {2020})}\BibitemShut
  {NoStop}%
\bibitem [{\citenamefont {Dzsotjan}\ \emph {et~al.}(2020)\citenamefont
  {Dzsotjan}, \citenamefont {Schmidt},\ and\ \citenamefont
  {Fleischhauer}}]{Dzsotjan2020}%
  \BibitemOpen
  \bibfield  {author} {\bibinfo {author} {\bibfnamefont {D.}~\bibnamefont
  {Dzsotjan}}, \bibinfo {author} {\bibfnamefont {R.}~\bibnamefont {Schmidt}},\
  and\ \bibinfo {author} {\bibfnamefont {M.}~\bibnamefont {Fleischhauer}},\
  }\bibfield  {title} {\bibinfo {title} {Dynamical {{Variational Approach}} to
  {{Bose Polarons}} at {{Finite Temperatures}}},\ }\href
  {https://doi.org/10.1103/PhysRevLett.124.223401} {\bibfield  {journal}
  {\bibinfo  {journal} {Physical Review Letters}\ }\textbf {\bibinfo {volume}
  {124}},\ \bibinfo {pages} {223401} (\bibinfo {year} {2020})}\BibitemShut
  {NoStop}%
\bibitem [{\citenamefont {Schmidt}\ and\ \citenamefont
  {Enss}(2022)}]{Schmidt2022}%
  \BibitemOpen
  \bibfield  {author} {\bibinfo {author} {\bibfnamefont {R.}~\bibnamefont
  {Schmidt}}\ and\ \bibinfo {author} {\bibfnamefont {T.}~\bibnamefont {Enss}},\
  }\bibfield  {title} {\bibinfo {title} {Self-stabilized {{Bose}} polarons},\
  }\href {https://doi.org/10.21468/SciPostPhys.13.3.054} {\bibfield  {journal}
  {\bibinfo  {journal} {SciPost Physics}\ }\textbf {\bibinfo {volume} {13}},\
  \bibinfo {pages} {054} (\bibinfo {year} {2022})}\BibitemShut {NoStop}%
\bibitem [{\citenamefont {Drescher}\ \emph {et~al.}(2021)\citenamefont
  {Drescher}, \citenamefont {Salmhofer},\ and\ \citenamefont
  {Enss}}]{Drescher2021}%
  \BibitemOpen
  \bibfield  {author} {\bibinfo {author} {\bibfnamefont {M.}~\bibnamefont
  {Drescher}}, \bibinfo {author} {\bibfnamefont {M.}~\bibnamefont
  {Salmhofer}},\ and\ \bibinfo {author} {\bibfnamefont {T.}~\bibnamefont
  {Enss}},\ }\bibfield  {title} {\bibinfo {title} {Quench {{Dynamics}} of the
  {{Ideal Bose Polaron}} at {{Zero}} and {{Nonzero Temperatures}}},\ }\href
  {https://doi.org/10.1103/PhysRevA.103.033317} {\bibfield  {journal} {\bibinfo
   {journal} {Physical Review A}\ }\textbf {\bibinfo {volume} {103}},\ \bibinfo
  {pages} {033317} (\bibinfo {year} {2021})}\BibitemShut {NoStop}%
\bibitem [{\citenamefont {Levinsen}\ \emph {et~al.}(2021)\citenamefont
  {Levinsen}, \citenamefont {Ardila}, \citenamefont {Yoshida},\ and\
  \citenamefont {Parish}}]{Levinsen2021}%
  \BibitemOpen
  \bibfield  {author} {\bibinfo {author} {\bibfnamefont {J.}~\bibnamefont
  {Levinsen}}, \bibinfo {author} {\bibfnamefont {L.~A.~P.}\ \bibnamefont
  {Ardila}}, \bibinfo {author} {\bibfnamefont {S.~M.}\ \bibnamefont
  {Yoshida}},\ and\ \bibinfo {author} {\bibfnamefont {M.~M.}\ \bibnamefont
  {Parish}},\ }\bibfield  {title} {\bibinfo {title} {Quantum {{Behavior}} of a
  {{Heavy Impurity Strongly Coupled}} to a {{Bose Gas}}},\ }\href
  {https://doi.org/10.1103/PhysRevLett.127.033401} {\bibfield  {journal}
  {\bibinfo  {journal} {Physical Review Letters}\ }\textbf {\bibinfo {volume}
  {127}},\ \bibinfo {pages} {033401} (\bibinfo {year} {2021})}\BibitemShut
  {NoStop}%
\bibitem [{\citenamefont {Rose}\ and\ \citenamefont
  {Schmidt}(2022)}]{Rose2022}%
  \BibitemOpen
  \bibfield  {author} {\bibinfo {author} {\bibfnamefont {F.}~\bibnamefont
  {Rose}}\ and\ \bibinfo {author} {\bibfnamefont {R.}~\bibnamefont {Schmidt}},\
  }\bibfield  {title} {\bibinfo {title} {Disorder in order: {{Localization}}
  without randomness in a cold-atom system},\ }\href
  {https://doi.org/10.1103/PhysRevA.105.013324} {\bibfield  {journal} {\bibinfo
   {journal} {Physical Review A}\ }\textbf {\bibinfo {volume} {105}},\ \bibinfo
  {pages} {013324} (\bibinfo {year} {2022})}\BibitemShut {NoStop}%
\bibitem [{\citenamefont {Enss}\ \emph {et~al.}(2022)\citenamefont {Enss},
  \citenamefont {Cuadra~Braatz},\ and\ \citenamefont {Gori}}]{Enss2022}%
  \BibitemOpen
  \bibfield  {author} {\bibinfo {author} {\bibfnamefont {T.}~\bibnamefont
  {Enss}}, \bibinfo {author} {\bibfnamefont {N.}~\bibnamefont
  {Cuadra~Braatz}},\ and\ \bibinfo {author} {\bibfnamefont {G.}~\bibnamefont
  {Gori}},\ }\bibfield  {title} {\bibinfo {title} {Complex scaling flows in the
  quench dynamics of interacting particles},\ }\href
  {https://doi.org/10.1103/PhysRevA.106.013308} {\bibfield  {journal} {\bibinfo
   {journal} {Physical Review A}\ }\textbf {\bibinfo {volume} {106}},\ \bibinfo
  {pages} {013308} (\bibinfo {year} {2022})},\ \Eprint
  {https://arxiv.org/abs/2203.06098} {arxiv:2203.06098} \BibitemShut {NoStop}%
\bibitem [{\citenamefont {Zinner}(2013)}]{Zinner2013}%
  \BibitemOpen
  \bibfield  {author} {\bibinfo {author} {\bibfnamefont {N.~T.}\ \bibnamefont
  {Zinner}},\ }\bibfield  {title} {\bibinfo {title} {Efimov states of heavy
  impurities in a {{Bose-Einstein}} condensate},\ }\href
  {https://doi.org/10.1209/0295-5075/101/60009} {\bibfield  {journal} {\bibinfo
   {journal} {EPL (Europhysics Letters)}\ }\textbf {\bibinfo {volume} {101}},\
  \bibinfo {pages} {60009} (\bibinfo {year} {2013})}\BibitemShut {NoStop}%
\bibitem [{\citenamefont {Naidon}(2018)}]{Naidon2018}%
  \BibitemOpen
  \bibfield  {author} {\bibinfo {author} {\bibfnamefont {P.}~\bibnamefont
  {Naidon}},\ }\bibfield  {title} {\bibinfo {title} {Two {{Impurities}} in a
  {{Bose}}\textendash{{Einstein Condensate}}: {{From Yukawa}} to {{Efimov
  Attracted Polarons}}},\ }\href {https://doi.org/10.7566/JPSJ.87.043002}
  {\bibfield  {journal} {\bibinfo  {journal} {Journal of the Physical Society
  of Japan}\ }\textbf {\bibinfo {volume} {87}},\ \bibinfo {pages} {043002}
  (\bibinfo {year} {2018})}\BibitemShut {NoStop}%
\bibitem [{\citenamefont {{Camacho-Guardian}}\ and\ \citenamefont
  {Bruun}(2018)}]{Camacho-Guardian2018}%
  \BibitemOpen
  \bibfield  {author} {\bibinfo {author} {\bibfnamefont {A.}~\bibnamefont
  {{Camacho-Guardian}}}\ and\ \bibinfo {author} {\bibfnamefont {G.~M.}\
  \bibnamefont {Bruun}},\ }\bibfield  {title} {\bibinfo {title} {Landau
  {{Effective Interaction}} between {{Quasiparticles}} in a {{Bose-Einstein
  Condensate}}},\ }\href {https://doi.org/10.1103/PhysRevX.8.031042} {\bibfield
   {journal} {\bibinfo  {journal} {Physical Review X}\ }\textbf {\bibinfo
  {volume} {8}},\ \bibinfo {pages} {031042} (\bibinfo {year}
  {2018})}\BibitemShut {NoStop}%
\bibitem [{\citenamefont {Panochko}\ and\ \citenamefont
  {Pastukhov}(2022)}]{Panochko2022}%
  \BibitemOpen
  \bibfield  {author} {\bibinfo {author} {\bibfnamefont {G.}~\bibnamefont
  {Panochko}}\ and\ \bibinfo {author} {\bibfnamefont {V.}~\bibnamefont
  {Pastukhov}},\ }\bibfield  {title} {\bibinfo {title} {Static {{Impurities}}
  in a {{Weakly Interacting Bose Gas}}},\ }\href
  {https://doi.org/10.3390/atoms10010019} {\bibfield  {journal} {\bibinfo
  {journal} {Atoms}\ }\textbf {\bibinfo {volume} {10}},\ \bibinfo {pages} {19}
  (\bibinfo {year} {2022})}\BibitemShut {NoStop}%
\bibitem [{\citenamefont {{Camacho-Guardian}}\ \emph
  {et~al.}(2018)\citenamefont {{Camacho-Guardian}}, \citenamefont
  {Pe{\~n}a~Ardila}, \citenamefont {Pohl},\ and\ \citenamefont
  {Bruun}}]{Camacho-Guardian2018a}%
  \BibitemOpen
  \bibfield  {author} {\bibinfo {author} {\bibfnamefont {A.}~\bibnamefont
  {{Camacho-Guardian}}}, \bibinfo {author} {\bibfnamefont {L.~A.}\ \bibnamefont
  {Pe{\~n}a~Ardila}}, \bibinfo {author} {\bibfnamefont {T.}~\bibnamefont
  {Pohl}},\ and\ \bibinfo {author} {\bibfnamefont {G.~M.}\ \bibnamefont
  {Bruun}},\ }\bibfield  {title} {\bibinfo {title} {Bipolarons in a
  {{Bose-Einstein Condensate}}},\ }\href
  {https://doi.org/10.1103/PhysRevLett.121.013401} {\bibfield  {journal}
  {\bibinfo  {journal} {Physical Review Letters}\ }\textbf {\bibinfo {volume}
  {121}},\ \bibinfo {pages} {013401} (\bibinfo {year} {2018})}\BibitemShut
  {NoStop}%
\bibitem [{\citenamefont {Panochko}\ and\ \citenamefont
  {Pastukhov}(2021)}]{Panochko2021}%
  \BibitemOpen
  \bibfield  {author} {\bibinfo {author} {\bibfnamefont {G.}~\bibnamefont
  {Panochko}}\ and\ \bibinfo {author} {\bibfnamefont {V.}~\bibnamefont
  {Pastukhov}},\ }\bibfield  {title} {\bibinfo {title} {Two- and three-body
  effective potentials between impurities in ideal {{BEC}}},\ }\href
  {https://doi.org/10.1088/1751-8121/abdbc5} {\bibfield  {journal} {\bibinfo
  {journal} {Journal of Physics A: Mathematical and Theoretical}\ }\textbf
  {\bibinfo {volume} {54}},\ \bibinfo {pages} {085001} (\bibinfo {year}
  {2021})}\BibitemShut {NoStop}%
\bibitem [{\citenamefont {Fujii}\ \emph {et~al.}(2022)\citenamefont {Fujii},
  \citenamefont {Hongo},\ and\ \citenamefont {Enss}}]{Fujii2022}%
  \BibitemOpen
  \bibfield  {author} {\bibinfo {author} {\bibfnamefont {K.}~\bibnamefont
  {Fujii}}, \bibinfo {author} {\bibfnamefont {M.}~\bibnamefont {Hongo}},\ and\
  \bibinfo {author} {\bibfnamefont {T.}~\bibnamefont {Enss}},\ }\bibfield
  {title} {\bibinfo {title} {Universal van der {{Waals Force}} between {{Heavy
  Polarons}} in {{Superfluids}}},\ }\href
  {https://doi.org/10.1103/PhysRevLett.129.233401} {\bibfield  {journal}
  {\bibinfo  {journal} {Physical Review Letters}\ }\textbf {\bibinfo {volume}
  {129}},\ \bibinfo {pages} {233401} (\bibinfo {year} {2022})}\BibitemShut
  {NoStop}%
\bibitem [{\citenamefont {Jager}\ and\ \citenamefont
  {Barnett}(2022)}]{Jager2022}%
  \BibitemOpen
  \bibfield  {author} {\bibinfo {author} {\bibfnamefont {J.}~\bibnamefont
  {Jager}}\ and\ \bibinfo {author} {\bibfnamefont {R.}~\bibnamefont
  {Barnett}},\ }\bibfield  {title} {\bibinfo {title} {The effect of
  boson\textendash boson interaction on the bipolaron formation},\ }\href
  {https://doi.org/10.1088/1367-2630/ac9804} {\bibfield  {journal} {\bibinfo
  {journal} {New Journal of Physics}\ }\textbf {\bibinfo {volume} {24}},\
  \bibinfo {pages} {103032} (\bibinfo {year} {2022})}\BibitemShut {NoStop}%
\bibitem [{\citenamefont {Ding}\ \emph {et~al.}(2022)\citenamefont {Ding},
  \citenamefont {Drewsen}, \citenamefont {Arlt},\ and\ \citenamefont
  {Bruun}}]{Ding2022}%
  \BibitemOpen
  \bibfield  {author} {\bibinfo {author} {\bibfnamefont {S.}~\bibnamefont
  {Ding}}, \bibinfo {author} {\bibfnamefont {M.}~\bibnamefont {Drewsen}},
  \bibinfo {author} {\bibfnamefont {J.~J.}\ \bibnamefont {Arlt}},\ and\
  \bibinfo {author} {\bibfnamefont {G.~M.}\ \bibnamefont {Bruun}},\ }\bibfield
  {title} {\bibinfo {title} {Mediated {{Interaction}} between {{Ions}} in
  {{Quantum Degenerate Gases}}},\ }\href
  {https://doi.org/10.1103/PhysRevLett.129.153401} {\bibfield  {journal}
  {\bibinfo  {journal} {Physical Review Letters}\ }\textbf {\bibinfo {volume}
  {129}},\ \bibinfo {pages} {153401} (\bibinfo {year} {2022})}\BibitemShut
  {NoStop}%
\bibitem [{\citenamefont {Astrakharchik}\ \emph {et~al.}(2023)\citenamefont
  {Astrakharchik}, \citenamefont {Ardila}, \citenamefont {Jachymski},\ and\
  \citenamefont {Negretti}}]{Astrakharchik2023}%
  \BibitemOpen
  \bibfield  {author} {\bibinfo {author} {\bibfnamefont {G.~E.}\ \bibnamefont
  {Astrakharchik}}, \bibinfo {author} {\bibfnamefont {L.~A.~P.}\ \bibnamefont
  {Ardila}}, \bibinfo {author} {\bibfnamefont {K.}~\bibnamefont {Jachymski}},\
  and\ \bibinfo {author} {\bibfnamefont {A.}~\bibnamefont {Negretti}},\
  }\bibfield  {title} {\bibinfo {title} {Many-body bound states and induced
  interactions of charged impurities in a bosonic bath},\ }\href
  {https://doi.org/10.1038/s41467-023-37153-0} {\bibfield  {journal} {\bibinfo
  {journal} {Nature Communications}\ }\textbf {\bibinfo {volume} {14}},\
  \bibinfo {pages} {1647} (\bibinfo {year} {2023})}\BibitemShut {NoStop}%
\bibitem [{\citenamefont {Bighin}\ \emph {et~al.}(2022)\citenamefont {Bighin},
  \citenamefont {Murthy}, \citenamefont {Defenu},\ and\ \citenamefont
  {Enss}}]{Bighin2022a}%
  \BibitemOpen
  \bibfield  {author} {\bibinfo {author} {\bibfnamefont {G.}~\bibnamefont
  {Bighin}}, \bibinfo {author} {\bibfnamefont {P.~A.}\ \bibnamefont {Murthy}},
  \bibinfo {author} {\bibfnamefont {N.}~\bibnamefont {Defenu}},\ and\ \bibinfo
  {author} {\bibfnamefont {T.}~\bibnamefont {Enss}},\ }\href
  {https://doi.org/10.48550/arXiv.2212.07419} {\bibinfo {title} {Resonantly
  enhanced superconductivity mediated by spinor condensates}} (\bibinfo {year}
  {2022}),\ \Eprint {https://arxiv.org/abs/2212.07419} {arxiv:2212.07419
  [cond-mat]} \BibitemShut {NoStop}%
\bibitem [{\citenamefont {Schecter}\ and\ \citenamefont
  {Kamenev}(2014)}]{Schecter2014}%
  \BibitemOpen
  \bibfield  {author} {\bibinfo {author} {\bibfnamefont {M.}~\bibnamefont
  {Schecter}}\ and\ \bibinfo {author} {\bibfnamefont {A.}~\bibnamefont
  {Kamenev}},\ }\bibfield  {title} {\bibinfo {title} {Phonon-{{Mediated Casimir
  Interaction}} between {{Mobile Impurities}} in {{One-Dimensional Quantum
  Liquids}}},\ }\href {https://doi.org/10.1103/PhysRevLett.112.155301}
  {\bibfield  {journal} {\bibinfo  {journal} {Physical Review Letters}\
  }\textbf {\bibinfo {volume} {112}},\ \bibinfo {pages} {155301} (\bibinfo
  {year} {2014})}\BibitemShut {NoStop}%
\bibitem [{\citenamefont {Dehkharghani}\ \emph {et~al.}(2018)\citenamefont
  {Dehkharghani}, \citenamefont {Volosniev},\ and\ \citenamefont
  {Zinner}}]{Dehkharghani2018}%
  \BibitemOpen
  \bibfield  {author} {\bibinfo {author} {\bibfnamefont {A.~S.}\ \bibnamefont
  {Dehkharghani}}, \bibinfo {author} {\bibfnamefont {A.~G.}\ \bibnamefont
  {Volosniev}},\ and\ \bibinfo {author} {\bibfnamefont {N.~T.}\ \bibnamefont
  {Zinner}},\ }\bibfield  {title} {\bibinfo {title} {Coalescence of {{Two
  Impurities}} in a {{Trapped One-dimensional Bose Gas}}},\ }\href
  {https://doi.org/10.1103/PhysRevLett.121.080405} {\bibfield  {journal}
  {\bibinfo  {journal} {Physical Review Letters}\ }\textbf {\bibinfo {volume}
  {121}},\ \bibinfo {pages} {080405} (\bibinfo {year} {2018})}\BibitemShut
  {NoStop}%
\bibitem [{\citenamefont {Charalambous}\ \emph {et~al.}(2019)\citenamefont
  {Charalambous}, \citenamefont {{Garcia-March}}, \citenamefont {Lampo},
  \citenamefont {Mehboud},\ and\ \citenamefont
  {Lewenstein}}]{Charalambous2019}%
  \BibitemOpen
  \bibfield  {author} {\bibinfo {author} {\bibfnamefont {C.}~\bibnamefont
  {Charalambous}}, \bibinfo {author} {\bibfnamefont {M.~A.}\ \bibnamefont
  {{Garcia-March}}}, \bibinfo {author} {\bibfnamefont {A.}~\bibnamefont
  {Lampo}}, \bibinfo {author} {\bibfnamefont {M.}~\bibnamefont {Mehboud}},\
  and\ \bibinfo {author} {\bibfnamefont {M.}~\bibnamefont {Lewenstein}},\
  }\bibfield  {title} {\bibinfo {title} {Two distinguishable impurities in
  {{BEC}}: Squeezing and entanglement of two {{Bose}} polarons},\ }\href
  {https://doi.org/10.21468/SciPostPhys.6.1.010} {\bibfield  {journal}
  {\bibinfo  {journal} {SciPost Physics}\ }\textbf {\bibinfo {volume} {6}},\
  \bibinfo {pages} {010} (\bibinfo {year} {2019})}\BibitemShut {NoStop}%
\bibitem [{\citenamefont {Mistakidis}\ \emph {et~al.}(2020)\citenamefont
  {Mistakidis}, \citenamefont {Volosniev},\ and\ \citenamefont
  {Schmelcher}}]{Mistakidis2020}%
  \BibitemOpen
  \bibfield  {author} {\bibinfo {author} {\bibfnamefont {S.~I.}\ \bibnamefont
  {Mistakidis}}, \bibinfo {author} {\bibfnamefont {A.~G.}\ \bibnamefont
  {Volosniev}},\ and\ \bibinfo {author} {\bibfnamefont {P.}~\bibnamefont
  {Schmelcher}},\ }\bibfield  {title} {\bibinfo {title} {Induced correlations
  between impurities in a one-dimensional quenched {{Bose}} gas},\ }\href
  {https://doi.org/10.1103/PhysRevResearch.2.023154} {\bibfield  {journal}
  {\bibinfo  {journal} {Physical Review Research}\ }\textbf {\bibinfo {volume}
  {2}},\ \bibinfo {pages} {023154} (\bibinfo {year} {2020})}\BibitemShut
  {NoStop}%
\bibitem [{\citenamefont {Will}\ \emph {et~al.}(2021)\citenamefont {Will},
  \citenamefont {Astrakharchik},\ and\ \citenamefont
  {Fleischhauer}}]{Will2021}%
  \BibitemOpen
  \bibfield  {author} {\bibinfo {author} {\bibfnamefont {M.}~\bibnamefont
  {Will}}, \bibinfo {author} {\bibfnamefont {G.~E.}\ \bibnamefont
  {Astrakharchik}},\ and\ \bibinfo {author} {\bibfnamefont {M.}~\bibnamefont
  {Fleischhauer}},\ }\bibfield  {title} {\bibinfo {title} {Polaron
  {{Interactions}} and {{Bipolarons}} in {{One-Dimensional Bose Gases}} in the
  {{Strong Coupling Regime}}},\ }\href
  {https://doi.org/10.1103/PhysRevLett.127.103401} {\bibfield  {journal}
  {\bibinfo  {journal} {Physical Review Letters}\ }\textbf {\bibinfo {volume}
  {127}},\ \bibinfo {pages} {103401} (\bibinfo {year} {2021})}\BibitemShut
  {NoStop}%
\bibitem [{\citenamefont {Petkovi{\'c}}\ and\ \citenamefont
  {Ristivojevic}(2022)}]{Petkovic2022}%
  \BibitemOpen
  \bibfield  {author} {\bibinfo {author} {\bibfnamefont {A.}~\bibnamefont
  {Petkovi{\'c}}}\ and\ \bibinfo {author} {\bibfnamefont {Z.}~\bibnamefont
  {Ristivojevic}},\ }\bibfield  {title} {\bibinfo {title} {Mediated interaction
  between polarons in a one-dimensional {{Bose}} gas},\ }\href
  {https://doi.org/10.1103/PhysRevA.105.L021303} {\bibfield  {journal}
  {\bibinfo  {journal} {Physical Review A}\ }\textbf {\bibinfo {volume}
  {105}},\ \bibinfo {pages} {L021303} (\bibinfo {year} {2022})},\ \Eprint
  {https://arxiv.org/abs/2103.08772} {arxiv:2103.08772} \BibitemShut {NoStop}%
\bibitem [{\citenamefont {Nishida}(2009)}]{Nishida2009}%
  \BibitemOpen
  \bibfield  {author} {\bibinfo {author} {\bibfnamefont {Y.}~\bibnamefont
  {Nishida}},\ }\bibfield  {title} {\bibinfo {title} {Casimir interaction among
  heavy fermions in the {{BCS-BEC}} crossover},\ }\href
  {https://doi.org/10.1103/PhysRevA.79.013629} {\bibfield  {journal} {\bibinfo
  {journal} {Physical Review A}\ }\textbf {\bibinfo {volume} {79}},\ \bibinfo
  {pages} {013629} (\bibinfo {year} {2009})}\BibitemShut {NoStop}%
\bibitem [{\citenamefont {DeSalvo}\ \emph {et~al.}(2019)\citenamefont
  {DeSalvo}, \citenamefont {Patel}, \citenamefont {Cai},\ and\ \citenamefont
  {Chin}}]{DeSalvo2019}%
  \BibitemOpen
  \bibfield  {author} {\bibinfo {author} {\bibfnamefont {B.~J.}\ \bibnamefont
  {DeSalvo}}, \bibinfo {author} {\bibfnamefont {K.}~\bibnamefont {Patel}},
  \bibinfo {author} {\bibfnamefont {G.}~\bibnamefont {Cai}},\ and\ \bibinfo
  {author} {\bibfnamefont {C.}~\bibnamefont {Chin}},\ }\bibfield  {title}
  {\bibinfo {title} {Observation of fermion-mediated interactions between
  bosonic atoms},\ }\href {https://doi.org/10.1038/s41586-019-1055-0}
  {\bibfield  {journal} {\bibinfo  {journal} {Nature}\ }\textbf {\bibinfo
  {volume} {568}},\ \bibinfo {pages} {61} (\bibinfo {year} {2019})}\BibitemShut
  {NoStop}%
\bibitem [{\citenamefont {Edri}\ \emph {et~al.}(2020)\citenamefont {Edri},
  \citenamefont {Raz}, \citenamefont {Matzliah}, \citenamefont {Davidson},\
  and\ \citenamefont {Ozeri}}]{Edri2020}%
  \BibitemOpen
  \bibfield  {author} {\bibinfo {author} {\bibfnamefont {H.}~\bibnamefont
  {Edri}}, \bibinfo {author} {\bibfnamefont {B.}~\bibnamefont {Raz}}, \bibinfo
  {author} {\bibfnamefont {N.}~\bibnamefont {Matzliah}}, \bibinfo {author}
  {\bibfnamefont {N.}~\bibnamefont {Davidson}},\ and\ \bibinfo {author}
  {\bibfnamefont {R.}~\bibnamefont {Ozeri}},\ }\bibfield  {title} {\bibinfo
  {title} {Observation of {{Spin-Spin Fermion-Mediated Interactions}} between
  {{Ultracold Bosons}}},\ }\href
  {https://doi.org/10.1103/PhysRevLett.124.163401} {\bibfield  {journal}
  {\bibinfo  {journal} {Physical Review Letters}\ }\textbf {\bibinfo {volume}
  {124}},\ \bibinfo {pages} {163401} (\bibinfo {year} {2020})}\BibitemShut
  {NoStop}%
\bibitem [{\citenamefont {Enss}\ \emph {et~al.}(2020)\citenamefont {Enss},
  \citenamefont {Tran}, \citenamefont {Rautenberg}, \citenamefont {Gerken},
  \citenamefont {Lippi}, \citenamefont {Drescher}, \citenamefont {Zhu},
  \citenamefont {Weidem{\"u}ller},\ and\ \citenamefont {Salmhofer}}]{Enss2020}%
  \BibitemOpen
  \bibfield  {author} {\bibinfo {author} {\bibfnamefont {T.}~\bibnamefont
  {Enss}}, \bibinfo {author} {\bibfnamefont {B.}~\bibnamefont {Tran}}, \bibinfo
  {author} {\bibfnamefont {M.}~\bibnamefont {Rautenberg}}, \bibinfo {author}
  {\bibfnamefont {M.}~\bibnamefont {Gerken}}, \bibinfo {author} {\bibfnamefont
  {E.}~\bibnamefont {Lippi}}, \bibinfo {author} {\bibfnamefont
  {M.}~\bibnamefont {Drescher}}, \bibinfo {author} {\bibfnamefont
  {B.}~\bibnamefont {Zhu}}, \bibinfo {author} {\bibfnamefont {M.}~\bibnamefont
  {Weidem{\"u}ller}},\ and\ \bibinfo {author} {\bibfnamefont {M.}~\bibnamefont
  {Salmhofer}},\ }\bibfield  {title} {\bibinfo {title} {Scattering of two heavy
  {{Fermi}} polarons: {{Resonances}} and quasibound states},\ }\href
  {https://doi.org/10.1103/PhysRevA.102.063321} {\bibfield  {journal} {\bibinfo
   {journal} {Physical Review A}\ }\textbf {\bibinfo {volume} {102}},\ \bibinfo
  {pages} {063321} (\bibinfo {year} {2020})}\BibitemShut {NoStop}%
\bibitem [{\citenamefont {Tran}\ \emph {et~al.}(2021)\citenamefont {Tran},
  \citenamefont {Rautenberg}, \citenamefont {Gerken}, \citenamefont {Lippi},
  \citenamefont {Zhu}, \citenamefont {Ulmanis}, \citenamefont {Drescher},
  \citenamefont {Salmhofer}, \citenamefont {Enss},\ and\ \citenamefont
  {Weidem{\"u}ller}}]{Tran2021}%
  \BibitemOpen
  \bibfield  {author} {\bibinfo {author} {\bibfnamefont {B.}~\bibnamefont
  {Tran}}, \bibinfo {author} {\bibfnamefont {M.}~\bibnamefont {Rautenberg}},
  \bibinfo {author} {\bibfnamefont {M.}~\bibnamefont {Gerken}}, \bibinfo
  {author} {\bibfnamefont {E.}~\bibnamefont {Lippi}}, \bibinfo {author}
  {\bibfnamefont {B.}~\bibnamefont {Zhu}}, \bibinfo {author} {\bibfnamefont
  {J.}~\bibnamefont {Ulmanis}}, \bibinfo {author} {\bibfnamefont
  {M.}~\bibnamefont {Drescher}}, \bibinfo {author} {\bibfnamefont
  {M.}~\bibnamefont {Salmhofer}}, \bibinfo {author} {\bibfnamefont
  {T.}~\bibnamefont {Enss}},\ and\ \bibinfo {author} {\bibfnamefont
  {M.}~\bibnamefont {Weidem{\"u}ller}},\ }\bibfield  {title} {\bibinfo {title}
  {Fermions {{Meet Two Bosons}}\textemdash the {{Heteronuclear Efimov Effect
  Revisited}}},\ }\href {https://doi.org/10.1007/s13538-020-00811-5} {\bibfield
   {journal} {\bibinfo  {journal} {Brazilian Journal of Physics}\ }\textbf
  {\bibinfo {volume} {51}},\ \bibinfo {pages} {316} (\bibinfo {year}
  {2021})}\BibitemShut {NoStop}%
\bibitem [{\citenamefont {Sun}\ \emph {et~al.}(2017)\citenamefont {Sun},
  \citenamefont {Zhai},\ and\ \citenamefont {Cui}}]{Sun2017}%
  \BibitemOpen
  \bibfield  {author} {\bibinfo {author} {\bibfnamefont {M.}~\bibnamefont
  {Sun}}, \bibinfo {author} {\bibfnamefont {H.}~\bibnamefont {Zhai}},\ and\
  \bibinfo {author} {\bibfnamefont {X.}~\bibnamefont {Cui}},\ }\bibfield
  {title} {\bibinfo {title} {Visualizing the {{Efimov Correlation}} in {{Bose
  Polarons}}},\ }\href {https://doi.org/10.1103/PhysRevLett.119.013401}
  {\bibfield  {journal} {\bibinfo  {journal} {Physical Review Letters}\
  }\textbf {\bibinfo {volume} {119}},\ \bibinfo {pages} {013401} (\bibinfo
  {year} {2017})}\BibitemShut {NoStop}%
\bibitem [{\citenamefont {Massignan}\ \emph {et~al.}(2021)\citenamefont
  {Massignan}, \citenamefont {Yegovtsev},\ and\ \citenamefont
  {Gurarie}}]{Massignan2021}%
  \BibitemOpen
  \bibfield  {author} {\bibinfo {author} {\bibfnamefont {P.}~\bibnamefont
  {Massignan}}, \bibinfo {author} {\bibfnamefont {N.}~\bibnamefont
  {Yegovtsev}},\ and\ \bibinfo {author} {\bibfnamefont {V.}~\bibnamefont
  {Gurarie}},\ }\bibfield  {title} {\bibinfo {title} {Universal {{Aspects}} of
  a {{Strongly Interacting Impurity}} in a {{Dilute Bose Condensate}}},\ }\href
  {https://doi.org/10.1103/PhysRevLett.126.123403} {\bibfield  {journal}
  {\bibinfo  {journal} {Physical Review Letters}\ }\textbf {\bibinfo {volume}
  {126}},\ \bibinfo {pages} {123403} (\bibinfo {year} {2021})}\BibitemShut
  {NoStop}%
\end{thebibliography}
\end{document}